\documentclass{WileyMSP-template}

\begin{document}

\pagestyle{fancy}
\rhead{\includegraphics[width=2.5cm]{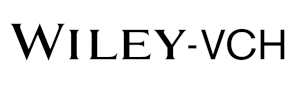}}

\title{Stochastic Dynamics of Domain Wall on a Racetrack: Impact of Line-Edge Roughness}

\maketitle


\author{Anton V. Hlushchenko}
\author{Oksana L. Andrieieva}
\author{Mykhailo I. Bratchenko}
\author{Andriy M. Styervoyedov}
\author{Kostyantyn I. Polozhiy}
\author{Aleksei V. Chechkin*}



\begin{affiliations}

Dr. Anton V. Hlushchenko, Dr. Oksana L. Andrieieva, Dr. Mykhailo I. Bratchenko\\
National Science Center "Kharkiv Institute of Physics and Technology"\\
61108 Kharkiv, Ukraine

Dr. Andriy M. Styervoyedov, Dr. Kostyantyn I. Polozhiy\\
Max Planck Institute of Microstructure Physics\\
06120 Halle, Germany

Prof. Dr. Aleksei V. Chechkin\\
National Science Center "Kharkiv Institute of Physics and Technology"\\
61108 Kharkiv, Ukraine\\
Max Planck Institute of Microstructure Physics\\
06120 Halle, Germany\\
Faculty of Pure and Applied Mathematics, Wrocław University of Science and Technology\\
50-370 Wrocław, Poland\\
Email Address: achechkin@mpi-halle.mpg.de

\end{affiliations}


\keywords{Racetrack, Domain Wall, Line-Edge Roughness, Stochastic pinning, Probabilistic Computing}

\begin{abstract}

We investigate the impact of line-edge roughness on current-driven domain wall dynamics in ferromagnetic racetracks. Modeling the edge disorder as a spatially correlated Ornstein-Uhlenbeck process, we demonstrate that even minimal experimentally relevant roughness induces pronounced stochastic pinning of domain walls. Notably, this stochasticity of the current-driven motion arises purely from spatial disorder, even in the absence of thermal fluctuations. The probability of a domain wall to reach a given position exhibits a robust sigmoidal dependence on the applied current, reflecting an effective distribution of depinning thresholds. At the same time, the underlying dynamics is highly nontrivial: the mean velocity exhibits a nonlinear dependence on both time and current, while the mean-square displacement exhibits a ballistic regime at short times followed by saturation due to trapping at pinning sites. These results demonstrate that line-edge roughness provides a controllable source of stochasticity and enables p-bit-like functionality in racetrack systems, offering a pathway toward hardware implementations of probabilistic and neuromorphic computing.

\end{abstract}

\section{Introduction}

Magnetic racetrack structures based on current-driven domain wall motion are a promising platform for non-volatile memory and logic devices \cite{Parkin_2008,Parkin2015,Parkin_2020}. In such systems, information is encoded in the spatial positions of domain walls, while spin-polarized currents are used to controllably manipulate their motion along the racetrack. The efficiency of this current-driven motion is highly sensitive to material parameters and interface properties \cite{Guan2021}, suggesting that even small variations can significantly affect domain wall dynamics. The dynamics of the domain wall on racetracks is inherently influenced by pinning \cite{Jiang2010,Gorchon_PhysRevLett_2014,Jeudy_PhysRevB_2018}, material disorder \cite{Reichhardt_RevModPhys_2022,Kaappa2024}, as well as thermal and current-induced fluctuations \cite{Tserkovnyak_shot_2005,Chudnovskiy_shot_2014,Hlushchenko2026}. These different sources of stochasticity manifest themselves in experimental observations of stochastic domain wall dynamics \cite{JaeChun_Science_2024,Ishibashi_SciAdv_2024}.

At the same time, stochasticity is increasingly recognized as a useful resource for computation. Probabilistic and unconventional computing paradigms have emerged as promising approaches for solving complex optimization and inference problems beyond conventional deterministic CMOS-based architectures \cite{Alaghi2013,Mohseni2022,Finocchio2024}. Their fundamental building block, the probabilistic bit (p-bit), is a stochastic unit that fluctuates between binary states with a tunable probability \cite{Camsari2017,Camsari2019}. Such systems exploit intrinsic fluctuations to enable efficient hardware implementations of stochastic neural networks, Boltzmann machines, and related architectures \cite{Liu2021,Lei2026,Grollier2020}.

Physically, p-bits can be realized in nanoscale systems exhibiting bistable states with controllable stochastic switching. In particular, low-barrier nanomagnets provide a natural platform where fluctuations induce random switching between two states, and the time-averaged response follows a sigmoidal dependence on the input bias \cite{Camsari2017}. This behavior directly maps onto activation functions used in probabilistic computing frameworks. Recent experiments have demonstrated on-chip p-bit implementations based on stochastic magnetic tunnel junctions \cite{Daniel2024}, highlighting the potential of spintronic devices for probabilistic and neuromorphic computing.
More generally, stochastic domain wall dynamics in confined magnetic systems can provide a physical basis for p-bit-like behavior, where magnetic textures undergo probabilistic transitions between metastable states \cite{Grollier2020,Godinho2024}.

In particular, structural disorder in racetrack geometries based on ferromagnetic thin films plays an important role, as it gives rise to stochastic pinning and depinning processes while remaining compatible with nanoscale device engineering. Among various sources of structural disorder, line-edge roughness (LER) plays an important role, originating from fabrication-induced imperfections such as stochastic fluctuations in the photoresist arising during lithographic processing, as well as etching-induced irregularities \cite{Albert2012,Dutta2017,Hoang2017}. Experimental studies have shown that LER leads to spatially distributed pinning sites along the edges of the nanotrack \cite{OShea2013}, while statistical approaches reveal that fluctuations in the domain wall reflect the landscape of the underlying disorder \cite{Jordan_PhysRevB_2020}.

Typical LER values in advanced lithographic processes are on the order of a few nanometers ($\approx 2$-$5$ nm), corresponding to a root-mean-square roughness amplitude $\sigma \approx 0.5$-$2$ nm and a correlation length $\xi \sim 5$-$50$ nm \cite{Constantoudis_2004,Mack2010,Naulleau_2010,Levinson2025,FernandezHerrero_2022,Kizu_2023}. Despite being often regarded as a fabrication imperfection, such a disorder can fundamentally alter domain wall dynamics and can be viewed as a controllable source of disorder in nanoscale devices.
Although stochasticity in spintronics devices is often associated with thermal noise \cite{Lei2026}, structural disorder provides an alternative and controllable mechanism to induce stochastic behavior by creating spatially varying pinning landscapes that lead to intermittent and nontrivial domain wall dynamics.

In this context, we investigate domain wall dynamics in ferromagnetic racetracks in the presence of line-edge roughness, modeled as a spatially correlated Ornstein-Uhlenbeck process. We demonstrate that such disorder alone induces stochastic pinning of domain walls, even in the absence of thermal fluctuations. The probability of a domain wall to reach a given position exhibits a robust sigmoidal dependence on the applied current, reflecting an effective distribution of depinning thresholds. At the same time, the underlying dynamics is highly nontrivial: the mean-square displacement (MSD) exhibits a ballistic regime at short times followed by saturation due to trapping, while the mean velocity exhibits a nonlinear dependence on both time and current, reflecting the interplay between driving and disorder.

These results show that line-edge roughness can be exploited to engineer stochastic behavior in racetrack systems, giving rise to probabilistic domain wall dynamics and enabling p-bit-like functionality. They also provide insight into domain wall transport in realistic devices, where line-edge-roughness-induced pinning leads to stochasticity in domain wall motion and increased current thresholds.

\begin{figure*}
\centering
\includegraphics[width=\linewidth]{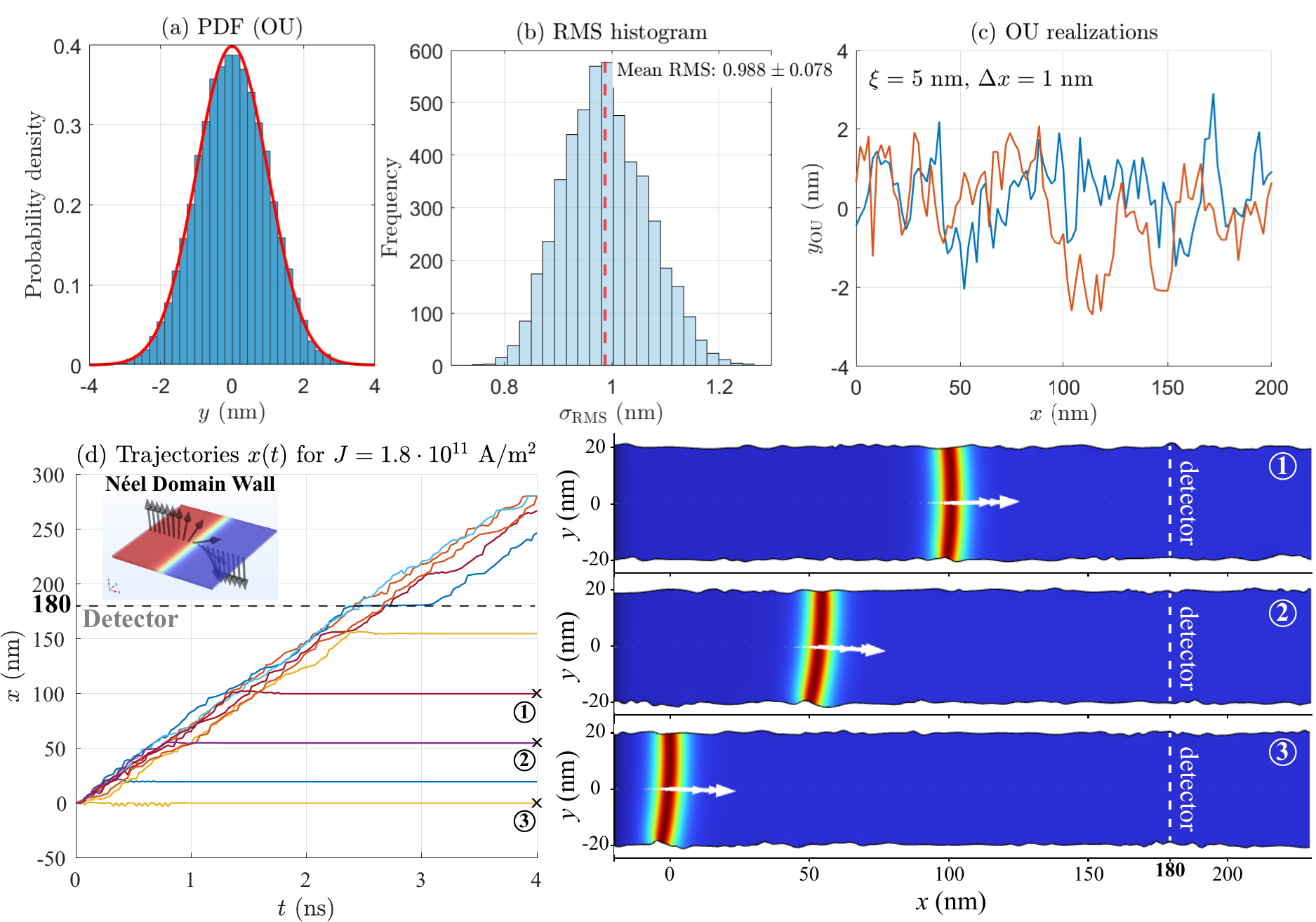}
\caption{\label{Displacements} (a) Probability density function (PDF) of the line-edge roughness modeled by an Ornstein-Uhlenbeck (OU) process, demonstrating a Gaussian distribution with a prescribed roughness amplitude $\sigma$.
(b) Distribution of the root-mean-square (RMS) roughness $\sigma_{\mathrm{RMS}}$ extracted from multiple realizations, highlighting statistical fluctuations arising from finite system size.
(c) Example realizations of the OU roughness profiles with correlation length $\xi = 5\,\mathrm{nm}$.
(d) Trajectories $x(t)$ of a domain wall in a racetrack with line-edge roughness modeled by the OU process at current density $J = 1.8 \times 10^{11}\,\mathrm{A/m^2}$, with $\sigma = 0.5\,\mathrm{nm}$ and $\xi = 5\,\mathrm{nm}$. The dashed line indicates the detector position. The panels (1)–(3) show representative domain wall configurations corresponding to the labeled trajectories.}
\end{figure*}

\section{Line-Edge Roughness on the Racetrack}
\begin{figure*}
\centering
\includegraphics[width=5.5cm]{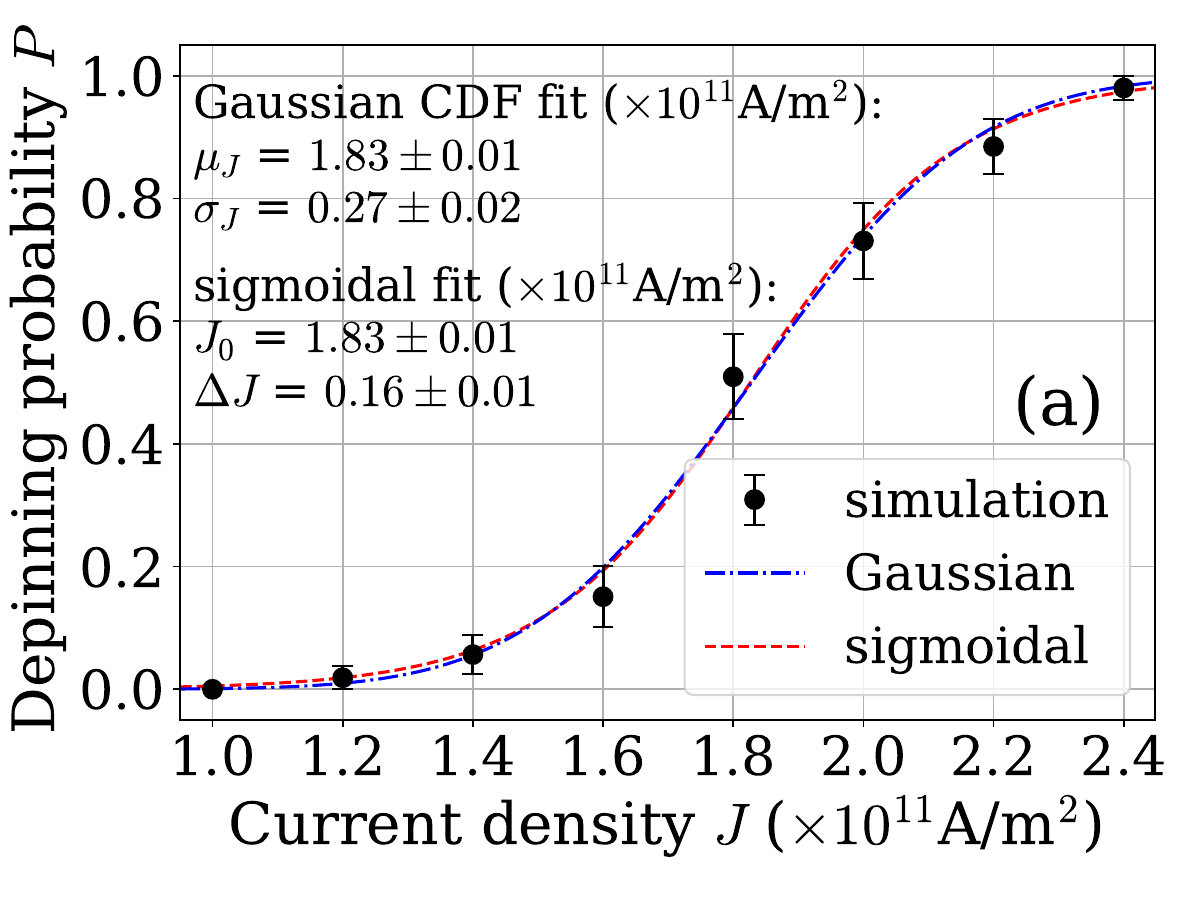}
\includegraphics[width=5.5cm]{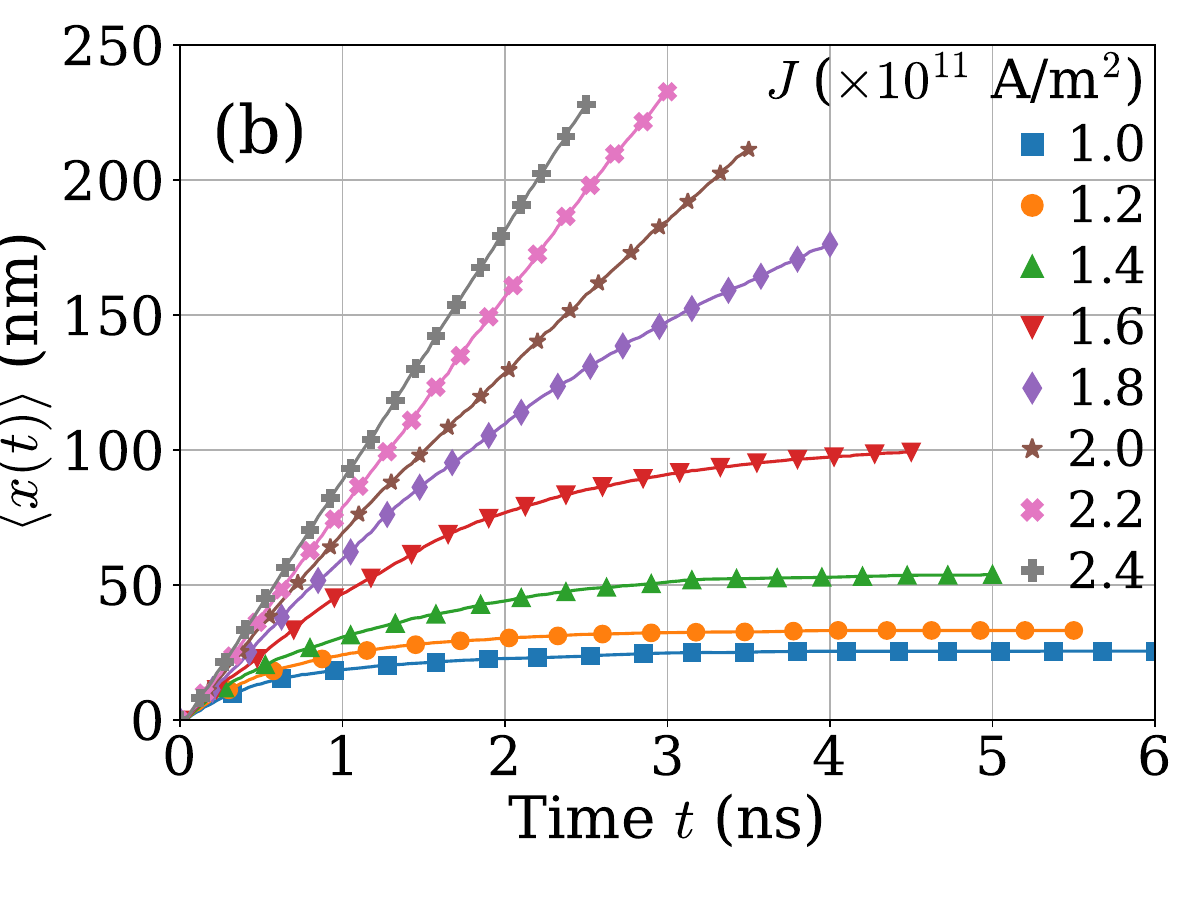}
\includegraphics[width=5.5cm]{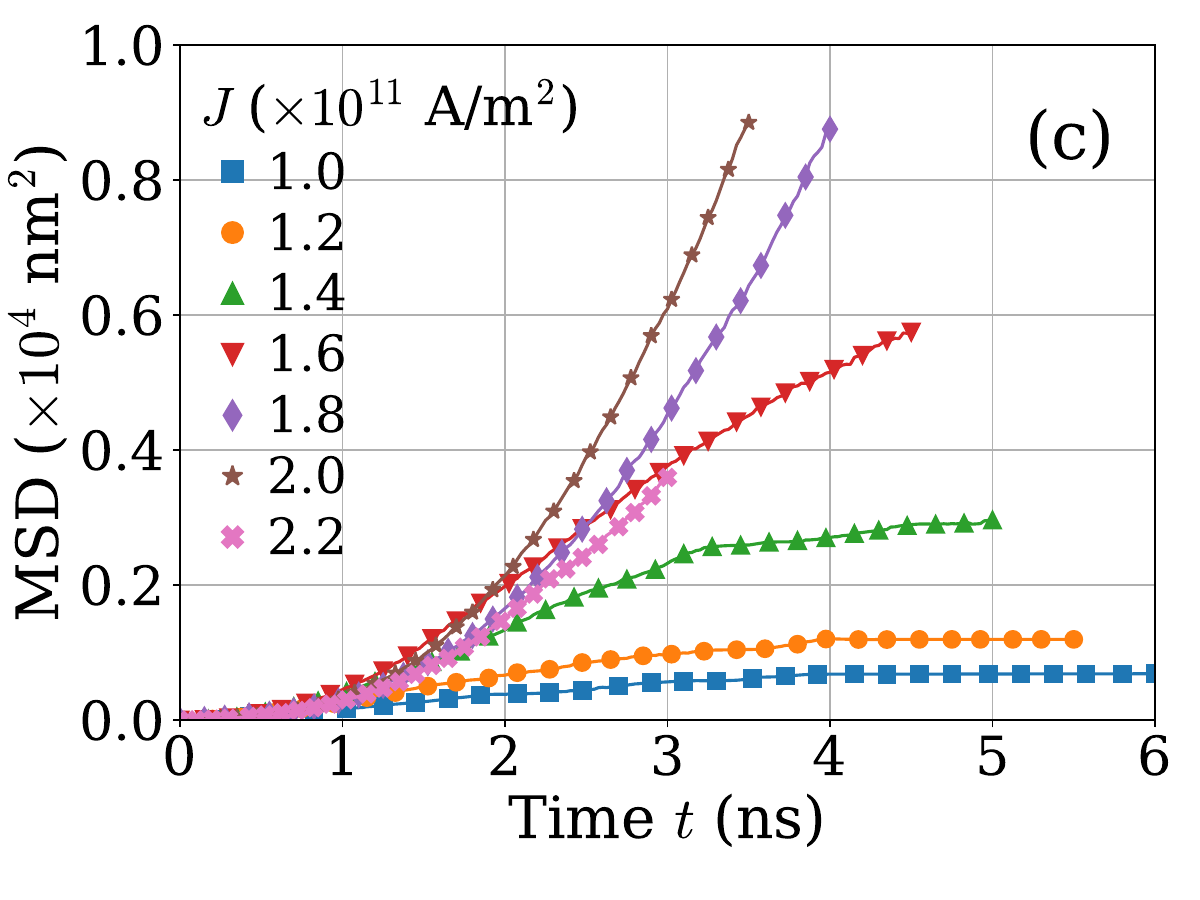}
\includegraphics[width=5.5cm]{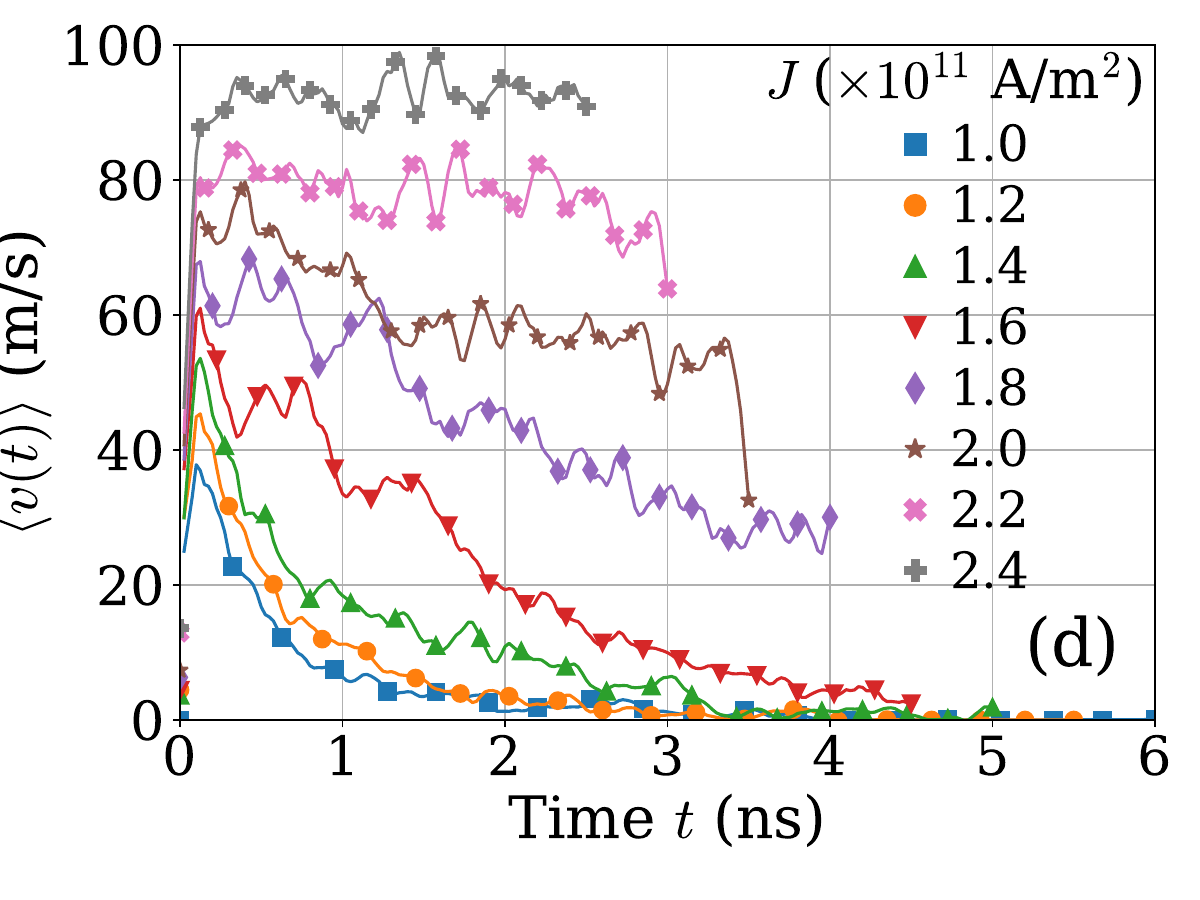}
\includegraphics[width=5.5cm]{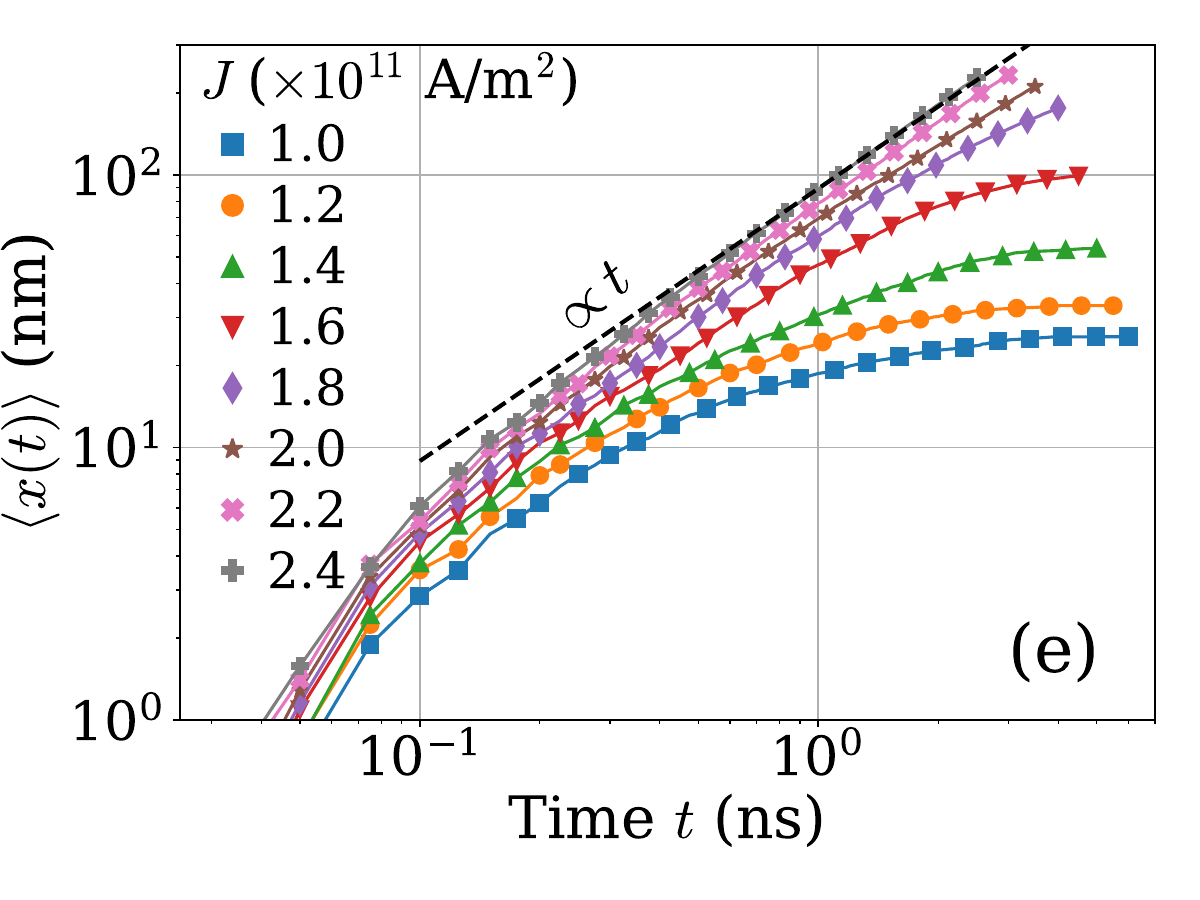}
\includegraphics[width=5.5cm]{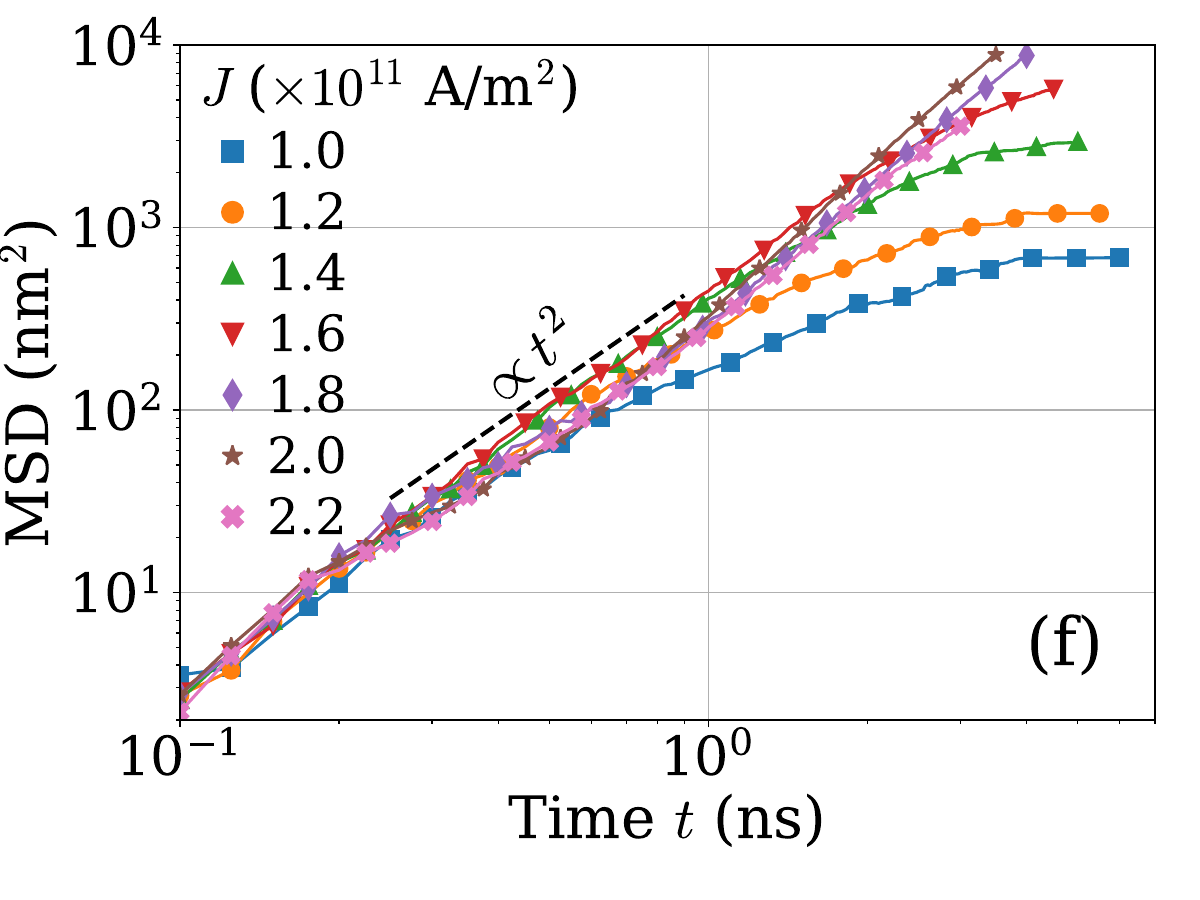}

\caption{\label{SurvivalProbability} (a) Depinning probability $P(J)$ as a function of current density $J$, obtained from simulations. Dashed and dash–dotted lines show fits to a Gaussian cumulative distribution and a logistic (sigmoidal) function, respectively.
(b) Mean domain wall position $\langle x(t) \rangle$ as a function of time for different current densities.
(c) Mean-square displacement as a function of time $t$, quantifying stochastic fluctuations around the mean trajectory.
(d) Time-dependent mean velocity $\langle v(t) \rangle$ for different current densities, demonstrating nonlinear behavior.
(e) Mean position $\langle x(t) \rangle$ in log–log scale. The dashed line indicates a linear dependence $\langle x(t) \rangle \propto t$ for comparison, highlighting the nonlinear character of the dynamics.
(f) MSD in log–log scale. The dashed line indicates superballistic scaling for comparison. At longer times, the curves approach a plateau, indicating saturation of the MSD.}
\end{figure*}

In realistic racetrack devices, fabrication-induced line-edge roughness (LER) induces spatial disorder along the nanotrack boundaries, which affects domain wall dynamics through stochastic pinning and depinning processes \cite{OShea2013}.
To model LER, we describe the edge fluctuations as a spatially correlated stochastic process. Specifically, we employ an Ornstein-Uhlenbeck (OU) process, which provides a minimal model of a Gaussian disorder with a finite correlation length. The OU process $y_{\mathrm{OU}}(x)$ is defined by the stochastic differential equation
\begin{equation}
\frac{d y_{\mathrm{OU}}}{dx} = -\frac{1}{\xi} y_{\mathrm{OU}}(x) + \eta(x),
\end{equation}
where $\xi$ is the length of the correlation and $\eta(x)$ is a Gaussian white noise term with zero mean and variance set by the roughness amplitude. This construction ensures that the resulting edge profile has a Gaussian probability distribution with standard deviation $\sigma$ and an exponentially decaying spatial correlation function,
\begin{equation}
\langle y_{\mathrm{OU}}(x) y_{\mathrm{OU}}(x') \rangle = \sigma^2 \exp\left(-\frac{|x - x'|}{\xi}\right).
\end{equation}

The geometry of the racetrack is then defined by two rough edges,
\begin{equation}
y_{\pm}(x) = \pm a + \sigma\, y_{\mathrm{OU}}(x),
\end{equation}
where $2a$ is the nominal width of the racetrack and $\sigma$ controls the amplitude of the edge fluctuations. This representation captures both the statistical properties of the experimentally observed LER and its spatial correlations.

In the following, we focus on the lower bound of experimentally relevant roughness parameters, choosing $\sigma = 0.5\,\mathrm{nm}$ and $\xi = 5\,\mathrm{nm}$, as discussed in the Introduction. This corresponds to a conservative scenario with minimal disorder strength. As will be shown below, even such weak line-edge roughness is sufficient to induce pronounced stochasticity in domain wall dynamics. Larger roughness amplitudes are expected to further enhance these effects without qualitatively altering the main features of the observed behavior.

The statistical properties of the roughness of the OU are illustrated in Fig.~\ref{Displacements}. Panel (a) shows the probability density function of $y_{\mathrm{OU}}$, which follows a Gaussian distribution with variance $\sigma^2$, confirming the expected behavior of the OU process. Panel (b) presents the distribution of the extracted root-mean-square roughness $\sigma_{\mathrm{RMS}}$ in 2000 realizations. Due to the finite size of the system $\sigma_{\mathrm{RMS}}$ exhibits fluctuations around the target value $\sigma$, as reflected by the spread of the histogram.
Representative realizations of the roughness profiles are shown in Fig.~\ref{Displacements}(c) for a correlation length $\xi = 5\,\mathrm{nm}$. These spatial correlations define the effective pinned landscape experienced by a propagating domain wall.

The impact of LER on the dynamics of the domain wall is illustrated in Fig.~\ref{Displacements}(d), which shows example trajectories $x(t)$ of a Néel domain wall driven by a current density $J = 1.8 \times 10^{11}\,\mathrm{A/m^2}$, obtained from micromagnetic simulations described in Sect.~IV. The trajectories display pronounced stochasticity, including variations in velocity, intermittent pinning, and, in some cases, the failure to reach the detector position. The corresponding domain wall configurations, shown in panels (1)-(3), highlight how local edge variations distort the wall and modify its propagation.
These results demonstrate that LER induces an effective spatially correlated disorder potential that governs domain wall motion. As a consequence, even in nominally identical devices, the propagation of domain walls becomes intrinsically stochastic, providing the physical basis for probabilistic functionality in racetrack systems.

\section{Stochastic Domain Wall Dynamics}

The presence of line-edge roughness gives rise to intrinsically stochastic domain wall motion, governed by the interplay of current-driven forces and pinning by spatially correlated disorder. In this section, we quantify this behavior in terms of depinning statistics, mean velocity, and mean squared displacement.

We begin by analyzing the probability of domain wall depinning as a function of the applied current density $J$. Figure~\ref{SurvivalProbability}(a) shows the depinning probability $P(J)$, where each data point is obtained from 50 independent realizations of line-edge roughness (see Sec.~IV for details), corresponding to distinct stochastic domain wall trajectories. The resulting dependence exhibits a smooth sigmoidal shape.

For comparison, fits using both a Gaussian cumulative distribution function (CDF) \cite{Lepadatu2009,Godinho2024} and a sigmoidal function \cite{Wang2025} are shown in Fig.~\ref{SurvivalProbability}(a) (dashed and dash-dotted lines, respectively). The two fits nearly coincide within the statistical uncertainty of the simulation data, indicating that the depinning probability is well described by a generic sigmoidal response.
For clarity, the Gaussian CDF and the sigmoidal function are given by
\begin{equation}
P(J) = \frac{1}{2} \left[1 + \mathrm{erf}\left(\frac{J - \mu_J}{\sigma_J \sqrt{2}}\right)\right],
\end{equation}
and
\begin{equation}
P(J) = \frac{1}{1 + \exp\left[-(J - J_0)/\Delta J\right]},
\end{equation}
respectively.

This behavior can be understood as arising from a distribution of depinning threshold currents associated with different disorder realizations. Each realization of the roughness landscape defines an effective pinning configuration and, consequently, a specific depinning threshold. The Gaussian fit yields a mean value $\mu_J = 1.83 \times 10^{11}\,\mathrm{A/m^2}$ and a standard deviation $\sigma_J = 0.27 \times 10^{11}\,\mathrm{A/m^2}$. The sigmoidal fit gives $J_0 = 1.83 \times 10^{11}\,\mathrm{A/m^2}$ and $\Delta J = 0.16 \times 10^{11}\,\mathrm{A/m^2}$. Both fits show good agreement not only in the mean value but also in the characteristic threshold current scale.
Although $\sigma_J$ and $\Delta J$ are not directly equivalent, they both characterize the width of the transition in different parameterizations. For a smooth crossover, the width is inversely related to the slope $dP/dJ$ at the inflection point: a steeper slope corresponds to a narrower transition region. For Gaussian- and sigmoid-like profiles, this leads to an approximate relation $\Delta J \approx \sigma_J \sqrt{2\pi}/4$.
At the same time, the fact that the depinning probability is well described by a sigmoidal-like function indicates that the system exhibits a generic input-output response characteristic of p-bits~\cite{Camsari2017}.

While the depinning probability obtained by averaging over the ensemble of the domain walls is well captured by this simple statistical description, the underlying domain wall dynamics is significantly more complex. This is illustrated in Fig.~\ref{SurvivalProbability}(b-d), which show the mean position $\langle x(t) \rangle$, mean-square displacement, and mean velocity $\langle v(t) \rangle$ for different current densities.
The mean position $\langle x(t) \rangle$ exhibits a nonlinear time dependence, deviating from simple linear propagation, as emphasized in the log-log representation in Fig.~\ref{SurvivalProbability}(e). The dashed line in Fig.~\ref{SurvivalProbability}(e) indicates a linear scaling $\langle x(t) \rangle \propto t$, highlighting the nonlinear character of the dynamics.

The current dependence of the mean velocity $\langle v(t) \rangle$, shown in Fig.~\ref{SurvivalProbability}(d), further reflects this complexity. At low current densities, domain wall motion is dominated by pinning, resulting in suppressed velocities and intermittent, creep-like dynamics. As the current increases, a crossover to a depinned regime occurs, characterized by a rapid increase in velocity. However, even in this regime, the motion remains strongly influenced by disorder, leading to fluctuations and deviations from a simple linear velocity-current relation.

Further insight into the stochastic nature of the dynamics is obtained from the MSD, defined as
\begin{equation}
\mathrm{MSD}(t) = \left\langle \big[x(t) - \langle x(t) \rangle\big]^2 \right\rangle,
\end{equation}
which quantifies fluctuations around the mean trajectory.

The MSD, shown in Fig.~\ref{SurvivalProbability}(c) and in log-log scale in Fig.~\ref{SurvivalProbability}(f), exhibits a rich temporal behavior that deviates from simple diffusive dynamics. At short times, the motion is close to ballistic, with $\mathrm{MSD}(t) \sim t^2$, as indicated by the dashed reference line in Fig.~\ref{SurvivalProbability}(f).
At longer times, the MSD curves approach a plateau, indicating saturation of fluctuations. This saturation reflects that domain wall trajectories eventually become trapped by pinning sites, suppressing further growth of the MSD.

These results highlight the difference between ensemble-averaged behavior and individual stochastic trajectories. Although the depinning probability follows a smooth and reproducible sigmoidal dependence on current, individual domain wall trajectories exhibit intermittency, pinning, and abrupt velocity variations.

\section{Conclusions}

In this work, we have investigated the impact of line-edge roughness on domain wall dynamics in ferromagnetic racetracks. By modeling the roughness as a spatially correlated Ornstein-Uhlenbeck process, we demonstrated that edge disorder alone induces stochastic pinning, leading to intrinsically probabilistic domain wall motion even in the absence of thermal fluctuations. The depinning probability exhibits a robust sigmoidal-like dependence on the applied current, reflecting an effective distribution of the depinning thresholds. At the same time, the underlying dynamics remains highly nontrivial. The mean velocity exhibits a nonlinear dependence on both time and current, reflecting transient dynamics and the influence of disorder-induced pinning. The mean-square displacement exhibits a ballistic regime at short times and subsequently approaches a plateau as a result of trapping at pinning sites.

These results highlight the coexistence of simple ensemble statistics and complex stochastic dynamics governed by spatially correlated disorder. They demonstrate that line-edge roughness provides a controllable source of stochasticity and enables p-bit-like functionality in racetrack systems, offering a pathway toward hardware implementations of probabilistic and neuromorphic computing. At the same time, these findings provide insight into domain wall dynamics in realistic racetrack devices, where disorder-induced pinning leads to variability in propagation and increased current thresholds for domain wall motion.


\section{Experimental Section}

\threesubsection{Micromagnetic Simulations}

The domain wall dynamics in the racetrack is modeled within the framework of micromagnetics by solving the Landau-Lifshitz-Gilbert equation

\begin{equation}
\frac{\partial \mathbf{m}}{\partial t} =
-\gamma \, \mathbf{m} \times \mathbf{H}_{\mathrm{eff}}
+ \alpha \, \mathbf{m} \times \frac{\partial \mathbf{m}}{\partial t}
+ \mathbf{T}_{\mathrm{SOT}}.
\end{equation}

\begin{equation}
\mathbf{H}_{\mathrm{eff}} =
\frac{2 A_\mathrm{ex}}{\mu_0 M_s} \nabla^2 \mathbf{m}
+ \frac{2 K_u}{\mu_0 M_s} m_z \mathbf{e}_z
+ \frac{2 D}{\mu_0 M_s}
\left( \nabla m_z - (\nabla \cdot \mathbf{m}) \mathbf{e}_z \right).
\end{equation}

\begin{equation}
\mathbf{T}_{\mathrm{SOT}} =
-\frac{\gamma \hbar P J}{2 e M_s \Delta_f}
\, \mathbf{m} \times \left( \mathbf{m} \times \mathbf{p} \right)
-\frac{\gamma \beta \hbar P J}{2 e M_s \Delta_f}
\, \mathbf{m} \times \mathbf{p}.
\end{equation}

Here, $\mathbf{m}$ denotes the unit magnetization vector, $\gamma$ the gyromagnetic ratio, and $\alpha$ the Gilbert damping constant. $A_\mathrm{ex}$ denotes the exchange stiffness, $K_u$ the perpendicular magnetic anisotropy constant, $D$ the interfacial Dzyaloshinskii-Moriya interaction constant, and $M_s$ the saturation magnetization. The value $\mu_0$ is the vacuum permeability, and $\mathbf{e}_z$ is the unit vector normal to the film plane.

The spin-orbit torque depends on the spin polarization $P$, current density $J$, reduced Planck constant $\hbar$, elementary charge $e$, and film thickness $\Delta_f$. The parameter  $\beta$ is the amplitude of the damping-like torque, and $\mathbf{p} = (\cos\psi, \sin\psi, 0)$ defines the spin-polarization direction.
No thermal fluctuations are included in the simulations, allowing us to isolate the effect of structural disorder on domain wall dynamics.

Micromagnetic simulations are performed using a finite-element framework implemented in COMSOL Multiphysics with the micromagnetic module \cite{comsolMicromag,micromagModule}. The racetrack geometry is defined as a thin ferromagnetic strip with line-edge roughness incorporated via spatial modulation of the racetrack boundaries, as described in Sec.~II. The magnetization dynamics is computed on a discretized mesh with a spatial resolution sufficient to resolve the internal structure of the domain wall.

The material parameters correspond to typical Pt/Co-based systems \cite{Guan2021,JaeChun_ACSNANO_2024}: exchange stiffness $A_\mathrm{ex} = 10^{-11}$~J/m, perpendicular anisotropy constant $K_u = 10^6$~J/m$^3$, saturation magnetization $M_s = 0.6 \times 10^{6}$~A/m, interfacial DMI constant $D = 0.5 \times 10^{-3}$~J/m$^2$, and Gilbert damping parameter $\alpha = 0.3$.
The racetrack has lateral dimensions of $400 \times 40$~nm$^2$ in the $(x,y)$ plane and a thickness $\Delta_f = 1$~nm. The simulations are performed using spatial discretization with a cell size of 1~nm.

The initial state consists of a Néel-type domain wall stabilized by interfacial DMI. The domain wall is driven along the racetrack by the applied current and its position $x(t)$ is extracted as a function of time. A total of 50 independent simulations with different realizations of line-edge roughness are performed to obtain statistically meaningful characteristics of the domain wall dynamics.


\medskip
\textbf{Acknowledgements} \par 
The authors acknowledge support from the BMFTR project 01DK24006 PLASMA-SPIN-ENERGY. AVH, OLA and MIB also acknowledge the National Research Foundation of Ukraine, Project No. 2025.07/0086 (Excellent Science in Ukraine 2026–2028).

\medskip
\bibliographystyle{MSP}
\bibliography{Stochastic}

@inproceedings{Mack2010,
  author    = {Chris A. Mack},
  title     = {Line-edge roughness and the ultimate limits of lithography},
  booktitle = {Proc. of SPIE},
  volume    = {7639},
  year      = {2010},
  doi       = {10.1117/12.838400}
}

@article{Levinson2025,
  author  = {Harry J. Levinson},
  title   = {Challenges and limits to patterning using extreme ultraviolet lithography},
  journal = {J. Micro/Nanopatterning Mater. Metrol.},
  volume  = {24},
  number  = {1},
  year    = {2025},
  doi     = {10.1117/1.JMM.24.1.011005}
}

@article{Naulleau_2010,
    author = {Naulleau, Patrick P. and Gallatin, Gregg M.},
    title = {Effect of resist on the transfer of line-edge roughness spatial metrics from mask to wafer},
    journal = {J. Vac. Sci. Technol., B	},
    volume = {28},
    number = {6},
    pages = {1259-1266},
    year = {2010},
    month = {11},
    issn = {2166-2746},
    doi = {10.1116/1.3509437},
    url = {https://doi.org/10.1116/1.3509437},
}

@article{Constantoudis_2004,
    author = {Constantoudis, V. and Patsis, G. P. and Leunissen, L. H. A. and Gogolides, E.},
    title = {Line edge roughness and critical dimension variation: Fractal characterization and comparison using model functions},
    journal = {J. Vac. Sci. Technol. B},
    volume = {22},
    number = {4},
    pages = {1974-1981},
    year = {2004},
    month = {08},
    issn = {1071-1023},
    doi = {10.1116/1.1776561},
    url = {https://doi.org/10.1116/1.1776561},
}

@article{Kizu_2023,
  author  = {Kizu, Ryosuke and Misumi, Ichiko and Hirai, Akiko and Gonda, Satoshi and Takahashi, Satoru},
  title   = {Developmental framework of line edge roughness reference standards for next-generation functional micro-/nanostructures},
  journal = {Precision Engineering},
  volume  = {83},
  pages   = {152--158},
  year    = {2023},
  doi     = {10.1016/j.precisioneng.2023.06.003}
}

@article{FernandezHerrero_2022,
  author  = {Fern{\'a}ndez Herrero, Anal{\'i}a and Scholze, Frank and Dai, Gaoliang and Soltwisch, Victor},
  title   = {Analysis of Line-Edge Roughness Using EUV Scatterometry},
  journal = {Nanomanufacturing and Metrology},
  volume  = {5},
  number  = {2},
  pages   = {149--158},
  year    = {2022},
  doi     = {10.1007/s41871-022-00126-w}
}

@article{Albert2012,
  author  = {Albert, Maximilian and Franchin, Matteo and Fischbacher, Thomas and Meier, Guido and Fangohr, Hans},
  title   = {Domain wall motion in perpendicular anisotropy nanowires with edge roughness},
  journal = {Journal of Physics: Condensed Matter},
  volume  = {24},
  number  = {2},
  pages   = {024219},
  year    = {2012},
  doi     = {10.1088/0953-8984/24/2/024219}
}

@article{Dutta2017,
  author  = {Dutta, Sumit and Siddiqui, Saima A. and Currivan-Incorvia, Jean Anne and Ross, Caroline A. and Baldo, Marc A.},
  title   = {The Spatial Resolution Limit for an Individual Domain Wall in Magnetic Nanowires},
  journal = {Nano Letters},
  volume  = {17},
  number  = {9},
  pages   = {5869--5874},
  year    = {2017},
  doi     = {10.1021/acs.nanolett.7b03199}
}

@article{Hoang2017,
  author  = {Hoang, Duc-Quang and Tran, Minh-Tung and Cao, Xuan-Huu and Ngo, Duc-The},
  title   = {The effect of edge roughness of magnetic nanowires on the degree of asymmetry in transverse domain walls},
  journal = {RSC Advances},
  volume  = {7},
  number  = {78},
  pages   = {49188--49193},
  year    = {2017},
  doi     = {10.1039/C7RA08104A}
}

@article{Camsari2017,
  author  = {Camsari, Kerem Yunus and Faria, Rafatul and Sutton, Brian M. and Datta, Supriyo},
  title   = {Stochastic $p$-Bits for Invertible Logic},
  journal = {Physical Review X},
  volume  = {7},
  number  = {3},
  pages   = {031014},
  year    = {2017},
  doi     = {10.1103/PhysRevX.7.031014}
}

@article{Camsari2019,
  author  = {Camsari, Kerem Y. and Sutton, Brian M. and Datta, Supriyo},
  title   = {p-bits for probabilistic spin logic},
  journal = {Applied Physics Reviews},
  volume  = {6},
  number  = {1},
  pages   = {011305},
  year    = {2019},
  doi     = {10.1063/1.5055860}
}

@article{Daniel2024,
  author  = {Daniel, John and Sun, Zheng and Zhang, Xuejian and Tan, Yuanqiu and Dilley, Neil and Chen, Zhihong and Appenzeller, Joerg},
  title   = {Experimental demonstration of an on-chip p-bit core based on stochastic magnetic tunnel junctions and 2D MoS$_2$ transistors},
  journal = {Nature Communications},
  volume  = {15},
  number  = {1},
  pages   = {4098},
  year    = {2024},
  doi     = {10.1038/s41467-024-48152-0}
}

@article{Godinho2024,
  author  = {Godinho, J. and Rout, P. K. and Salikhov, R. and Hellwig, O. and {\L}obacz, Z. and Otxoa, R. M. and Olejn{\'i}k, K. and Jungwirth, T. and Wunderlich, J.},
  title   = {Antiferromagnetic domain wall memory with neuromorphic functionality},
  journal = {npj Spintronics},
  volume  = {2},
  number  = {1},
  pages   = {39},
  year    = {2024},
  doi     = {10.1038/s44306-024-00027-2}
}

@article{Wang2025,
  author  = {Wang, Yadi and Chen, Bin and Gao, Wenping and Ye, Biying and Niu, Chang and Wang, Wenbin and Zhu, Yinyan and Yu, Weichao and Guo, Hangwen and Shen, Jian},
  title   = {Superior probabilistic computing using an operationally stable probabilistic bit constructed by a manganite nanowire},
  journal = {National Science Review},
  volume  = {12},
  number  = {3},
  pages   = {nwae338},
  year    = {2025},
  doi     = {10.1093/nsr/nwae338}
}

@article{OShea2013,
  author = {O'Shea, K. J. and Tracey, J. and Bramsiepe, S. and Stamps, R. L.},
  title = {Probing nanowire edge roughness using an extended magnetic domain wall},
  journal = {Applied Physics Letters},
  volume = {102},
  number = {6},
  pages = {062409},
  year = {2013},
  doi = {10.1063/1.4792314}
}

@article{Jordan_PhysRevB_2020,
  title = {Statistically meaningful measure of domain-wall roughness in magnetic thin films},
  author = {Jord\'an, D. and Albornoz, L. J. and Gorchon, J. and Lambert, C. H. and Salahuddin, S. and Bokor, J. and Curiale, J. and Bustingorry, S.},
  journal = {Phys. Rev. B},
  volume = {101},
  issue = {18},
  pages = {184431},
  numpages = {11},
  year = {2020},
  month = {May},
  publisher = {American Physical Society},
  doi = {10.1103/PhysRevB.101.184431},
  url = {https://link.aps.org/doi/10.1103/PhysRevB.101.184431}
}

@article{Kaappa2024,
  author = {Kaappa, Sami and Santa-aho, Suvi and Honkanen, Mari and Vippola, Minnamari and Laurson, Lasse},
  title = {Magnetic domain walls interacting with dislocations in micromagnetic simulations},
  journal = {Communications Materials},
  volume = {5},
  pages = {256},
  year = {2024},
  doi = {10.1038/s43246-024-00697-9}
}

@article{Hlushchenko2026,
  author = {Hlushchenko, Anton V. and Bratchenko, Mykhailo I. and Chechkin, Aleksei V.},
  title = {Stochastic Dynamics of Skyrmions on a Racetrack: Impact of Equilibrium and Nonequilibrium Noise},
  journal = {arXiv preprint arXiv:2511.20287},
  year = {2026}
}

@Article{Jiang2010,
author={Jiang, Xin
and Thomas, Luc
and Moriya, Rai
and Hayashi, Masamitsu
and Bergman, Bastiaan
and Rettner, Charles
and Parkin, Stuart S.P.},
title={Enhanced stochasticity of domain wall motion in magnetic racetracks due to dynamic pinning},
journal={Nat. Commun.},
year={2010},
month={Jun},
day={15},
volume={1},
number={1},
pages={25},
issn={2041-1723},
doi={10.1038/ncomms1024},
url={https://doi.org/10.1038/ncomms1024}
}

@article{Parkin_2008,
author = {Stuart S. P. Parkin  and Masamitsu Hayashi  and Luc Thomas},
title = {Magnetic Domain-Wall Racetrack Memory},
journal = {Science},
volume = {320},
number = {5873},
pages = {190-194},
year = {2008},
doi = {10.1126/science.1145799},
URL = {https://www.science.org/doi/abs/10.1126/science.1145799}
}

@Article{Parkin2015,
author={Parkin, Stuart
and Yang, See-Hun},
title={Memory on the racetrack},
journal={Nat. Nanotechnol.},
year={2015},
month={Mar},
day={01},
volume={10},
number={3},
pages={195-198},
issn={1748-3395},
doi={10.1038/nnano.2015.41},
url={https://doi.org/10.1038/nnano.2015.41}
}

@ARTICLE{Parkin_2020,
  author={Bl{\"a}sing, Robin and Khan, Asif Ali and Filippou, Panagiotis Ch and Garg, Chirag and Hameed, Fazal and Castrillon, Jeronimo and Parkin, Stuart SP},
  journal={Proc. IEEE}, 
  title={Magnetic Racetrack Memory: From Physics to the Cusp of Applications Within a Decade}, 
  year={2020},
  volume={108},
  number={8},
  pages={1303-1321},
  url={https://doi.org/10.1109/JPROC.2020.2975719},
  doi={10.1109/JPROC.2020.2975719}}

@article{JaeChun_ACSNANO_2024,
author = {Jeon, Jae-Chun and Migliorini, Andrea and Fischer, Lukas and Yoon, Jiho and Parkin, Stuart S. P.},
title = {Dynamic Manipulation of Chiral Domain Wall Spacing for Advanced Spintronic Memory and Logic Devices},
journal = {ACS Nano},
volume = {18},
number = {22},
pages = {14507-14513},
year = {2024},
doi = {10.1021/acsnano.4c02024},
URL = {https://doi.org/10.1021/acsnano.4c02024}
}

@article{JaeChun_Science_2024,
author = {Jae-Chun Jeon  and Andrea Migliorini  and Jiho Yoon  and Jaewoo Jeong  and Stuart S. P. Parkin },
title = {Multicore memristor from electrically readable nanoscopic racetracks},
journal = {Science},
volume = {386},
number = {6719},
pages = {315-322},
year = {2024},
doi = {10.1126/science.adh3419},
URL = {https://www.science.org/doi/abs/10.1126/science.adh3419}
}

@article{Chudnovskiy_shot_2014,
author = {Chudnovskiy, A. and H{\"u}bner, Ch. and Baxevanis, B. and Pfannkuche, D.},
title = {Spin switching: From quantum to quasiclassical approach},
journal = {Phys. Status Solidi B},
volume = {251},
number = {9},
pages = {1764-1776},
doi = {https://doi.org/10.1002/pssb.201350225},
url = {https://onlinelibrary.wiley.com/doi/abs/10.1002/pssb.201350225},
year = {2014}
}

@article{Tserkovnyak_shot_2005,
  title = {Magnetization Noise in Magnetoelectronic Nanostructures},
  author = {Foros, J\o{}rn and Brataas, Arne and Tserkovnyak, Yaroslav and Bauer, Gerrit E. W.},
  journal = {Phys. Rev. Lett.},
  volume = {95},
  issue = {1},
  pages = {016601},
  numpages = {4},
  year = {2005},
  month = {Jun},
  publisher = {American Physical Society},
  doi = {10.1103/PhysRevLett.95.016601},
  url = {https://link.aps.org/doi/10.1103/PhysRevLett.95.016601}
}

@article{Gorchon_PhysRevLett_2014,
  title = {Pinning-Dependent Field-Driven Domain Wall Dynamics and Thermal Scaling in an Ultrathin $\mathrm{Pt}/\mathrm{Co}/\mathrm{Pt}$ Magnetic Film},
  author = {Gorchon, J. and Bustingorry, S. and Ferr\'e, J. and Jeudy, V. and Kolton, A. B. and Giamarchi, T.},
  journal = {Phys. Rev. Lett.},
  volume = {113},
  issue = {2},
  pages = {027205},
  numpages = {5},
  year = {2014},
  month = {Jul},
  publisher = {American Physical Society},
  doi = {10.1103/PhysRevLett.113.027205},
  url = {https://link.aps.org/doi/10.1103/PhysRevLett.113.027205}
}

@article{Jeudy_PhysRevB_2018,
  title = {Pinning of domain walls in thin ferromagnetic films},
  author = {Jeudy, V. and D\'{\i}az Pardo, R. and Savero Torres, W. and Bustingorry, S. and Kolton, A. B.},
  journal = {Phys. Rev. B},
  volume = {98},
  issue = {5},
  pages = {054406},
  numpages = {11},
  year = {2018},
  month = {Aug},
  publisher = {American Physical Society},
  doi = {10.1103/PhysRevB.98.054406},
  url = {https://link.aps.org/doi/10.1103/PhysRevB.98.054406}
}

@article{Reichhardt_RevModPhys_2022,
  title = {Statics and dynamics of skyrmions interacting with disorder and nanostructures},
  author = {Reichhardt, C. and Reichhardt, C. J. O. and Milo\ifmmode \check{s}\else \v{s}\fi{}evi\ifmmode \acute{c}\else \'{c}\fi{}, M. V.},
  journal = {Rev. Mod. Phys.},
  volume = {94},
  issue = {3},
  pages = {035005},
  numpages = {61},
  year = {2022},
  month = {Sep},
  publisher = {American Physical Society},
  doi = {10.1103/RevModPhys.94.035005},
  url = {https://link.aps.org/doi/10.1103/RevModPhys.94.035005}
}

@article{Ishibashi_SciAdv_2024,
author = {Mio Ishibashi  and Masashi Kawaguchi  and Yuki Hibino  and Kay Yakushiji  and Arata Tsukamoto  and Satoru Nakatsuji  and Masamitsu Hayashi},
title = {Decoding the magnetic bit positioning error in a ferrimagnetic racetrack},
journal = {Sci. Adv.},
volume = {10},
number = {43},
pages = {eadq0898},
year = {2024},
doi = {10.1126/sciadv.adq0898},
URL = {https://www.science.org/doi/abs/10.1126/sciadv.adq0898},
}

@article{Alaghi2013,
  author = {Alaghi, Armin and Hayes, John P.},
  title = {Survey of Stochastic Computing},
  journal = {ACM Transactions on Embedded Computing Systems},
  volume = {12},
  number = {2s},
  pages = {92},
  year = {2013},
  doi = {10.1145/2465787.2465794}
}

@article{Liu2021,
  author = {Liu, Yidong and Liu, Siting and Wang, Yanzhi and Lombardi, Fabrizio and Han, Jie},
  title = {A Survey of Stochastic Computing Neural Networks for Machine Learning Applications},
  journal = {IEEE Transactions on Neural Networks and Learning Systems},
  volume = {32},
  number = {7},
  pages = {2809--2824},
  year = {2021},
  doi = {10.1109/TNNLS.2020.3009047}
}

@article{Lei2026,
  author = {Lei, Kun and Wang, Xinrui and Zhou, Yiqi and Liu, Bohan and Wang, Kaiyou},
  title = {Spin–Orbit Torque MRAM: Fundamentals, Device Engineering, and Applications in Probabilistic Computing},
  journal = {Advanced Quantum Technologies},
  volume = {9},
  number = {3},
  pages = {e00886},
  year = {2026},
  doi = {10.1002/qute.202500886}
}

@article{Grollier2020,
  author = {Grollier, J. and Querlioz, D. and Camsari, K. Y. and Everschor-Sitte, K. and Fukami, S. and Stiles, M. D.},
  title = {Neuromorphic spintronics},
  journal = {Nature Electronics},
  volume = {3},
  number = {7},
  pages = {360--370},
  year = {2020},
  doi = {10.1038/s41928-019-0360-9}
}

@article{Finocchio2024,
  author = {Finocchio, Giovanni and Incorvia, Jean Anne C. and Friedman, Joseph S. and Yang, Qu and Giordano, Anna and Grollier, Julie and Yang, Hyunsoo and Ciubotaru, Florin and Chumak, Andrii V. and Naeemi, Azad J. and Cotofana, Sorin D. and Tomasello, Riccardo and Panagopoulos, Christos and Carpentieri, Mario and Lin, Peng and Pan, Gang and Yang, J. Joshua and Todri-Sanial, Aida and Boschetto, Gabriele and Makasheva, Kremena and Sangwan, Vinod K. and Trivedi, Amit Ranjan and Hersam, Mark C. and Camsari, Kerem Y. and McMahon, Peter L. and Datta, Supriyo and Koiller, Belita and Aguilar, Gabriel H. and Temporao, Guilherme P. and Rodrigues, Davi R. and Sunada, Satoshi and Everschor-Sitte, Karin and Tatsumura, Kosuke and Goto, Hayato and Puliafito, Vito and Akerman, Johan and Takesue, Hiroki and Di Ventra, Massimiliano and Pershin, Yuriy V. and Mukhopadhyay, Saibal and Roy, Kaushik and Wang, I.-Ting and Kang, Wang and Zhu, Yao and Kaushik, Brajesh Kumar and Hasler, Jennifer and Ganguly, Samiran and Ghosh, Avik W. and Levy, William and Roychowdhury, Vwani and Bandyopadhyay, Supriyo},
  title = {Roadmap for unconventional computing with nanotechnology},
  journal = {Nano Futures},
  volume = {8},
  number = {1},
  pages = {012001},
  year = {2024},
  doi = {10.1088/2399-1984/ad299a}
}

@article{Mohseni2022,
  author = {Mohseni, Naeimeh and McMahon, Peter L. and Byrnes, Tim},
  title = {Ising machines as hardware solvers of combinatorial optimization problems},
  journal = {Nature Reviews Physics},
  volume = {4},
  number = {6},
  pages = {363--379},
  year = {2022},
  doi = {10.1038/s42254-022-00440-8}
}

@article{Guan2021,
  author = {Guan, Yicheng and Zhou, Xilin and Ma, Tianping and Bl\"asing, Robin and Deniz, Hakan and Yang, See-Hun and Parkin, Stuart S. P.},
  title = {Increased Efficiency of Current-Induced Motion of Chiral Domain Walls by Interface Engineering},
  journal = {Advanced Materials},
  volume = {33},
  number = {10},
  pages = {2007991},
  year = {2021},
  doi = {10.1002/adma.202007991}
}

@misc{comsolMicromag,
  author       = {Weichao Yu},
  title        = {Micromagnetics Module for COMSOL Multiphysics},
  howpublished = {\url{https://www.comsol.com/community/exchange/883/}},
  note         = {Version 2.13, Accessed: 2025-09-15}
}

@manual{micromagModule,
  author       = {Weichao Yu},
  title        = {Micromagnetics Module User's Guide (V1.33)},
  organization = {Fudan University},
  year         = {2020},
  howpublished = {\url{http://www.physics.fudan.edu.cn/tps/people/jxiao/micromagnetics-module-users.pdf}},
  note         = {Accessed: 2025-09-15}
}

@article{Lepadatu2009,
  title = {Dependence of Domain-Wall Depinning Threshold Current on Pinning Profile},
  author = {Lepadatu, S. and Vanhaverbeke, A. and Atkinson, D. and Allenspach, R. and Marrows, C. H.},
  journal = {Phys. Rev. Lett.},
  volume = {102},
  pages = {127203},
  year = {2009},
  doi = {10.1103/PhysRevLett.102.127203}
}

\end{document}